\documentclass[12pt]{article}\usepackage[hyperfootnotes=false]{hyperref}
   \usepackage{epsfig}
      \usepackage{amsmath}
  \usepackage{graphicx}
  \newlength{\abstractwidth}
  \renewcommand{\thefootnote}{\fnsymbol{footnote}}
  \renewcommand{\thanks}[1]{\footnote{#1}} 
  \newcommand{\starttext}{
  \setcounter{footnote}{0}
  \renewcommand{\thefootnote}{\arabic{footnote}}}
  \renewcommand{\theequation}{\thesection.\arabic{equation}}
  \newcommand{\be}{\begin{equation}}
  \newcommand{\bea}{\begin{eqnarray}}
  \newcommand{\eea}{\end{eqnarray}}
  \newcommand{\beq}{\begin{equation}}
  \newcommand{\ee}{\end{equation}}
  \newcommand{\eeq}{\end{equation}}

  \def\ba{\begin{eqnarray}}
  \def\ea{\end{eqnarray}}

  \def\12{{1 \over 2}}

  \def\d{\partial}

  \def\simleq{\; \raise0.3ex\hbox{$<$\kern-0.75em
      \raise-1.1ex\hbox{$\sim$}}\; }
   \def\simgeq{\; \raise0.3ex\hbox{$>$\kern-0.75em
      \raise-1.1ex\hbox{$\sim$}}\; }

\def\cdl{Coleman De Luccia}

\def\ba{\bf{a}}

  \def\h3{{\cal{H}}_3}

\def\o3{\Omega_3}

\def\O2{\Omega_2}
\def\o{\omega}
 \def\d2{$dS_{2+1}$}
\def\21{$(2+1)$-dimensional}
\def\31{$(3+1)$-dimensional}
 \def\bi{\begin{itemize}}
  \def\ei{\end{itemize}}

\begin{document}
  \renewcommand{\theequation}{\thesection.\arabic{equation}}


  \begin{titlepage}
  \rightline{}
  \bigskip

  \bigskip\bigskip\bigskip\bigskip

    \centerline{\Large \bf {Is Eternal Inflation Past-Eternal?}}

        \bigskip

        \centerline{\Large \bf {And  What if It Is?}}
    \bigskip

  \bigskip \bigskip

  \bigskip\bigskip
  \bigskip\bigskip

  \begin{center}
  {{ Leonard Susskind}}
  \bigskip

\bigskip
Stanford Institute for Theoretical Physics and  Department of Physics, Stanford University\\
Stanford, CA 94305-4060, USA \\

\vspace{2cm}
  \end{center}

 \bigskip\bigskip
  \begin{abstract}


As a result of discussions with Bousso and Vilenkin I want to return to the question of whether the multiverse is past-eternal or if there was a beginning. Not surprisingly, given three people, there were three answers. However, the discussions have led to some common ground.

The  multiverse being past-eternal, or at least extremely old has content and potential phenomenological implications. I will discuss how the oldness of the multiverse is connected with recent speculations of Douglas.

 \medskip
  \noindent
  \end{abstract}

  \end{titlepage}
  \starttext \baselineskip=17.63pt \setcounter{footnote}{0}


\tableofcontents

\section{The Argument for an Eternal Past}

In \cite{Mithani:2012ii}, Mithani and Vilenkin argue that the universe could not be past-eternal and must have had a beginning  (a.k.a., initial condition or past singularity). Subsequently
in \cite{Susskind:2012yv} I argued that (in a sense to be described) an inflating universe which is  future-eternal  must also be past-eternal. In stating this conclusion I tried to choose my words carefully. For example:

\bigskip

\it ``..in any kind of inflating cosmology the odds strongly
(infinitely) favor the beginning to be so far in the past that it is effectively at minus infinity.
More precisely, given any T the probability is unity that the beginning was more than T
time-units ago." \rm

\bigskip

I did not claim that the concept of an initial condition is unnecessary. I don't know if it is or not. 
In this paper I will elaborate on the various arguments and counterarguments for an eternal past, including those of Mithani-Vilenkin  and of Bousso\footnote{I thank A. Vilenkin and R. Bousso for explaining their viewpoints in a very interesting 3-way email exchange.}.

As a preliminary consider  stable de Sitter space.
For generality we can consider a landscape  in which all vacua have positive cosmological constants.  We may assume an initial starting point at time $t=0$, for example an initial vacuum on the landscape. As explained in \cite{Harlow:2011az} and \cite{Susskind:2012pp} if we follow a causal patch it will pass through an infinite number of Boltzmann fluctuations that will sample all vacua. Most observer's will be freaks but even if we condition on normal observers, there will be an infinite number of them. Given any finite time $T,$  all but a finite number of them will occur later than that time.  Thus if they think they are typical, observers should bet,  with overwhelming confidence, that they cannot detect a beginning.

Let's turn to case of greatest interest: eternal inflation.
 For purposes of exposition I will use the tree-model of  \cite{Harlow:2011az} which I believe has all the relevant features of eternal inflation, and has the added advantage of being exactly solvable.

\begin{figure}[h]
\begin{center}
\includegraphics[scale=.3]{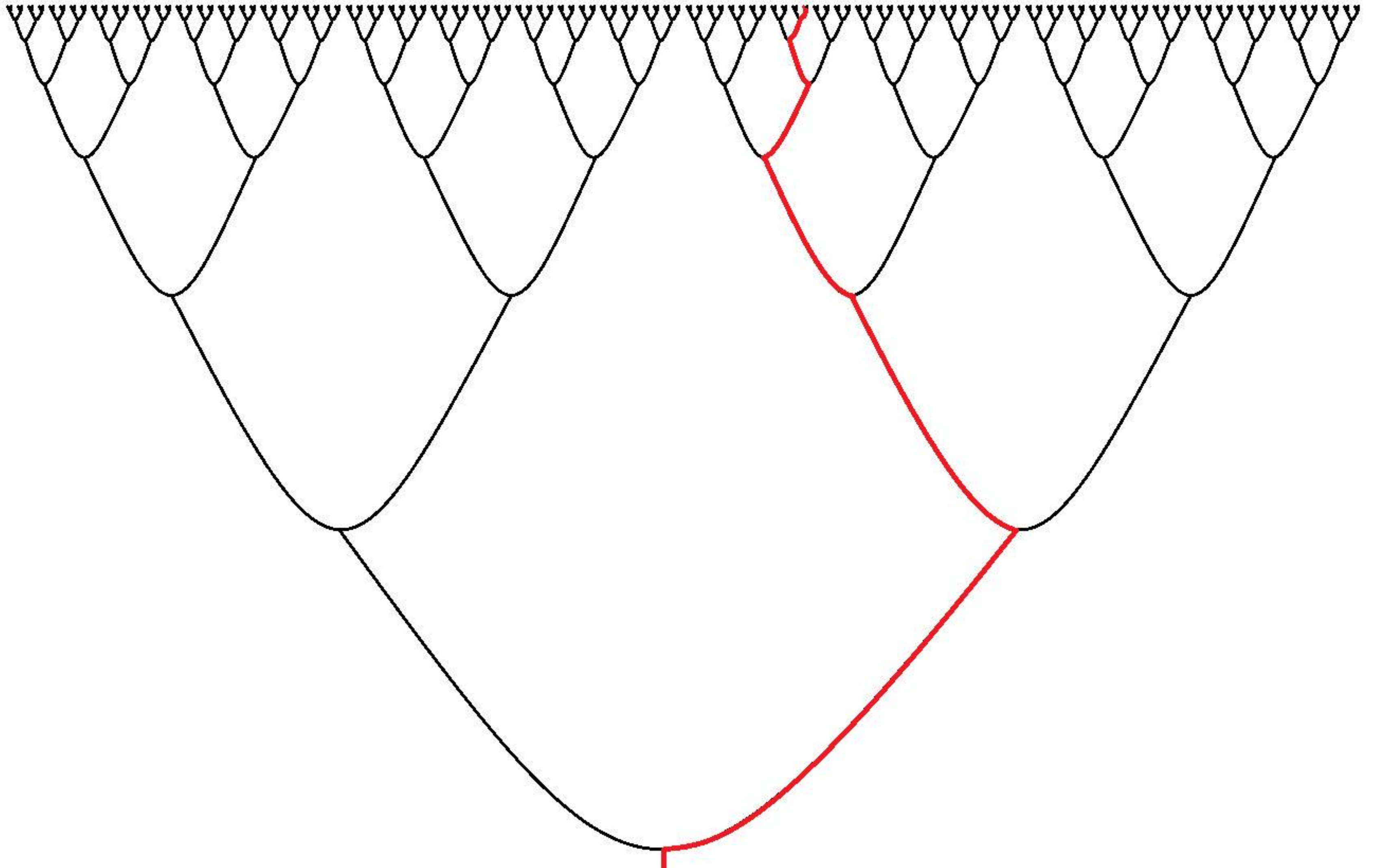}
\caption{The Multiverse According to the Tree Model. The red trajectory represents a single causal patch or observable history on an observer who ends on the future boundary of the tree. }
\label{f1}
\end{center}
\end{figure}

Each edge (link) of the tree represents a hubble time along the causal patch. It is labeled by an index $n$ called the color index. The color represents the  vacuum-type in the landscape (of positive energy  vacua). Associated with the color is an expansion rate $H_n$ and a Hawking-Gibbons entropy $S_n \sim \frac{1}{H_n^2}.$

Transitions from color $n$ to color $m$ occur  with rate $\gamma_{mn} \ll 1.$ Detailed balance requires the rates to satisfy
\be
\gamma_{mn} = M_{mn} e^{S_m}
\ee
where $M_{mn}$ is a symmetric matrix. Numerically transitions which increase the entropy (decrease $H$ ) are much more probable than transitions which decrease the entropy but both types are important in the long-time evolution of a causal patch.

In addition to de Sitter vacua there are terminal vacua. The details of the terminals will be unimportant other than the decay rates from the de Sitter to the terminal vacua. Note that the color-notation $m$ applies only to de Sitter vacua and not to terminals.
The rate $\gamma_n$ is the total rate for the vacuum $n$ to decay to terminals.

The integer-valued time along the tree is called $u.$ At a given time the average number of links (or causal patches) with vacuum-type $m$ satisfies the rate equation
\be
N_m(u+1) = (2-\gamma_m) N_m(u) -\sum_n \gamma_{nm}N_m(u) - \sum_n \gamma_{mn}N_n(u)
\label{global rate}
\ee

Note that if the rates to terminals are set to zero, the total number of links $\sum_n N(n)$ grows like $2^u.$  This simply represents the exponential growth of the multiverse with time. The decay to terminals makes the growth slightly slower but still exponential.

One can also define the fraction of vacua with a given color,
\be
P_m = \frac{N_m}{\sum_nN_n}.
\ee
The $P_m$ satisfy the rate equation
\be
P_m(u+1) = (1-\gamma_m) P_m(u) -\sum_n \gamma_{nm}P_m(u) - \sum_n \gamma_{mn}P_n(u)
\label{local rate}
\ee
If the decay to terminals is set to zero the sum of the $P_m$ is conserved. Otherwise it tends to zero, representing the  draining of probability  into the terminals.

The rate equations are linear and therefore the solutions can be written as sums over a complete set of eigenvectors $P^I$ in the form
\be
P_m(u) = \sum_I P^I_m \lambda_I^u
\ee
where the eigenvalues $\lambda_I$ satisfy
\be
    0< \lambda_I\leq1
\ee

The asymptotic $u\to \infty$  attractor is determined by the eigenvector with largest eigenvalue. In the limit the ratios of the $P_m$ are completely insensitive to any initial state such as an initial color.

There are two cases of interest: with and without terminals.
The solution to the rate equations without terminals exhibits the following features.

\bi

\item The largest eigenvalue is $\lambda_0 =1.$

\item The corresponding eigenvector is
\be
P^0_m =\frac{1}{z} e^{S_m }
\ee
where $z$ is a normalization factor. This attractor is just the equilibrium attractor with equal probability for every microstate. It corresponds to the Hartle Hawking no-boundary vacuum.

\item Correlation functions can be computed on the future boundary of the tree. The correlation functions exhibit a remarkable conformal symmetry entirely analogous to the conformal symmetry of the future boundary of de Sitter space.

\ei

The theory with terminals behaves differently. In this case:

\bi
\item The largest eigenvalue is strictly less than one. The deviation from unity is of order the decay rates to terminals and therefore very small.

\item    The largest eigenvalue and corresponing eigenvector are called $\lambda_D$ and $P^D_m.$ The notation $D$ means dominant. $P^D_m$ is not proportional to $e^{S_m }.$ It tends to be dominated by the longest-lived vacuum and neighboring vacua on the landscape. Neighboring means: can be reached by a small number of transitions. To my knowledge the concept of the dominant eigenvector first appeared in the work of Garriga, Schwartz-Perlov, Vilenkin, and S.Winitzki \cite{Garriga:2006hw}

    \item The conformal symmetry is broken. This is equivalent to the breaking of de Sitter symmetry. It picks out a preferred time-slicing which is the tree version of the so-called light-cone time.
\ei

Let us consider the root of the tree at $u=0$ to be the beginning. This is just a model  but it will serve the purpose. As a model of initial conditions we could specify the color of the root. There are various possibilities. One would be to choose the largest vacuum energy, and therefore the least entropy, for the root. The other extreme is also possible and suggested by the Hartle-Hawking state, namely the smallest positive vacuum energy. In either case, or for anything in between,  the asymptotic late-time population of vacua is completely insensitive to the starting point. It is entirely determined by the dominant eigenvector.

In making the argument for an eternal past I will use some assumptions that are more or less conventional. The question of a measure will be settled by assuming a cutoff procedure in which a cutoff time is introduced. Detailed predictions depend on the choice of time variable but not for the questions addressed in this paper. To be definite I will choose the natural time on the tree, namely $u.$ The cutoff time is called $u_0.$  Probabilities are calculated by first computing frequencies in the cutoff theory and then letting $u_0 \to \infty.$

Consider the number of observers $N_{obs}$ that are within time $U$ of the root of the tree. Because the population of every vacuum increases exponentially the answer is dominated by the observers at time $U$. If there are no terminals the number is proportional to $2^U.$ If there are terminals then
\be
N_{obs}(U) =( 2\lambda_D)^U
\label{U}
\ee

Next consider the total number of observers in the entire multiverse up to the cutoff time $u_0.$
 That number is
\be
N_{obs}(u_0) =( 2\lambda_D)^{u_0}
\label{u0}
\ee
Clearly the fraction of observers who are no further from the root that time $U$ is
\be
f = ( 2\lambda_D)^{(U-u_0 ) }.
\ee

Since this goes to zero exponentially,  the usual frequentist interpretation  implies that the probability to be within time $U$ of the root is zero. Moreover that is true for any $U.$
To put it another way, for all but a vanishing fraction of observers, with probability unity the root is in the infinite past.

That concludes the argument of \cite{Susskind:2012yv}. It applies to any of the usual cutoff procedures and measure proposals with the exception of the causal patch measure. I will return to that point.

\section{The Argument for a Beginning}

Let me now review the counterargument for a beginning and discuss its validity \cite{Mithani:2012ii}.
The argument is based on the observation in \cite{Borde:2001nh} that an inflating geometry must be past-incomplete. To see this in a simple context consider
 de Sitter space in flat slicing. The metric is given by
\be
ds^2 = -dt^2 +e^{2Ht}dx^i dx^i.
\label{ds}
\ee

\begin{figure}[h]
\begin{center}
\includegraphics[scale=.4]{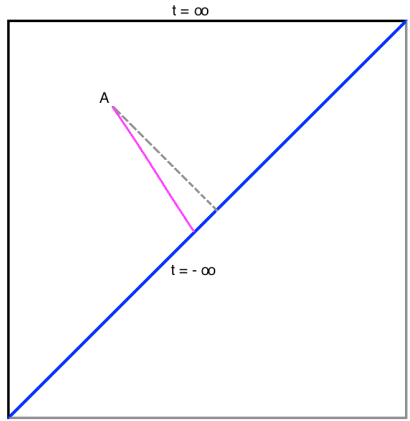}
\caption{The Penrose diagram for de Sitter space. The portion of de Sitter space covered by the coordinates in \ref{ds} is the upper left triangle of the diagram and the blue diagonal represents $t= -\infty.$ The dotted grey line is light-like and the pink line is time-like. The proper distance from $A$ to $t= -\infty$ vanishes along the light-like line and is finite along the time-like line.}
\label{f2}
\end{center}
\end{figure}

\begin{figure}[h]
\begin{center}
\includegraphics[scale=.3]{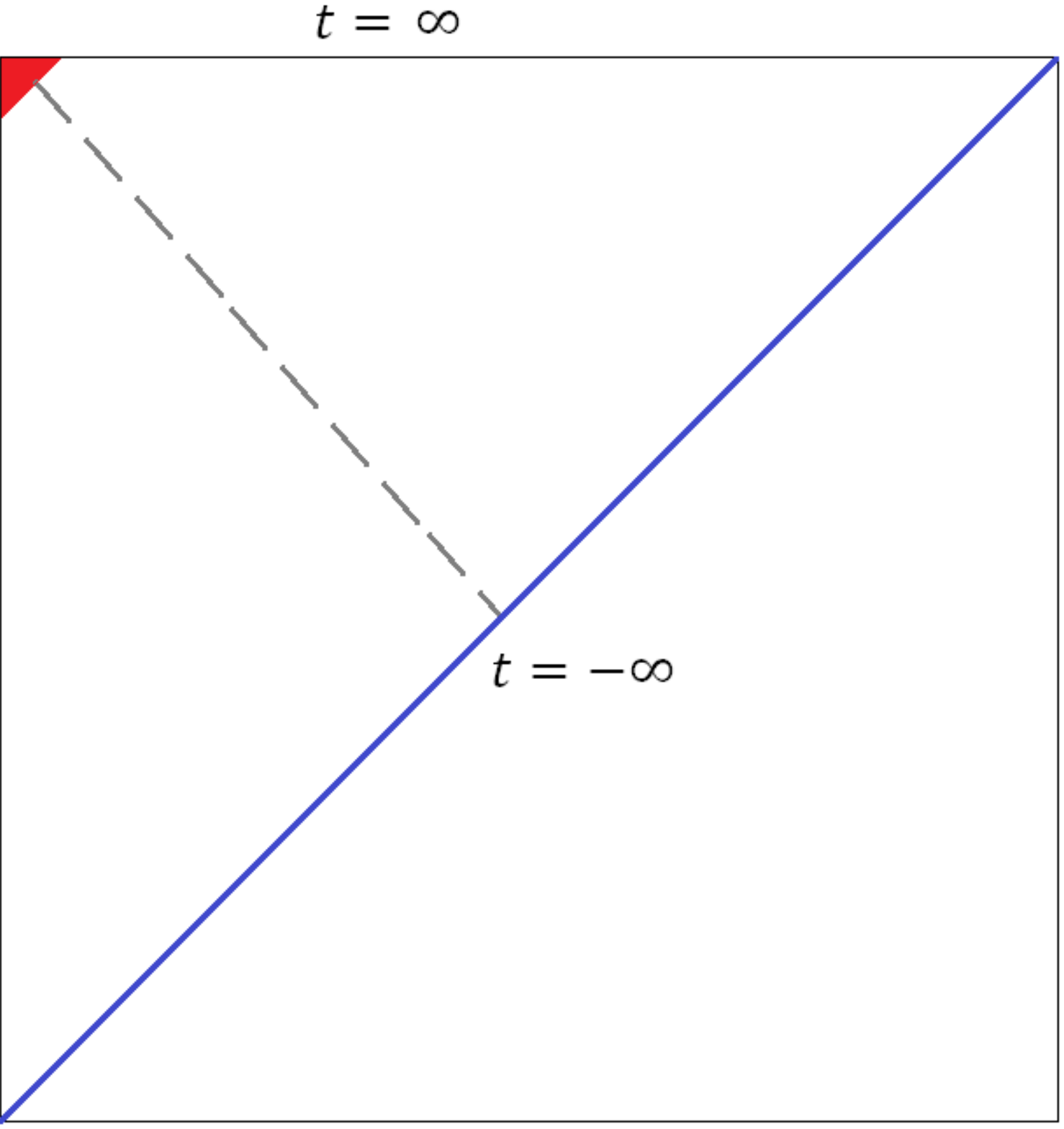}
\caption{The Guth Vilenkin argument. The blue line is time $-\infty$  in the flat slicing of de Sitter space. The red patch is a nucleated bubble. The grey broken line is a short time-like geodesic extending from the bubble all the ways back to $-\infty.$ }
\label{f3}
\end{center}
\end{figure}

Naively, the surface $t=-\infty$ appears to be the infinite past. But by drawing a Penrose diagram in Figure \ref{f2} one can see that this is a misleading identification. From any point $A$ at positive time, the surface $t=-\infty$ can be reached by a finite-length time-like geodesic. In fact that geodesic can be made to have arbitrarily small proper time. This would seem to say that points at finite times are in direct causal contact with,  and can receive strong detectable signals from, $t=-\infty.$
For example a wave-like disturbance on the surface $t=-\infty$ would be easily detectable.   Stated in this way, it is not surprising that an observer in the red bubble in Figure \ref{f3} can detect the initial condition at $t=-\infty$ as if it were \ a finite time-like distance away.

This argument seems compelling until one considers that the overwhelming majority of observers exist at asymptotically late time. It is clear from the figure that if the red bubble is pushed up to very late time---squeezed into the upper left corner---any signal from the blue line will be extremely red shifted. Whatever the initial conditions are at $t=-\infty,$ it becomes harder and harder to resolve the signal as the red bubble recedes to the future. Thus with a finite resolving power,  observers in the red bubble can only resolve features on the initial condition surface if those observers are not too late. Since the overwhelming majority of observers are later than any preassigned time, the probability to detect such features vanishes.

\begin{figure}[h]
\begin{center}
\includegraphics[scale=.3]{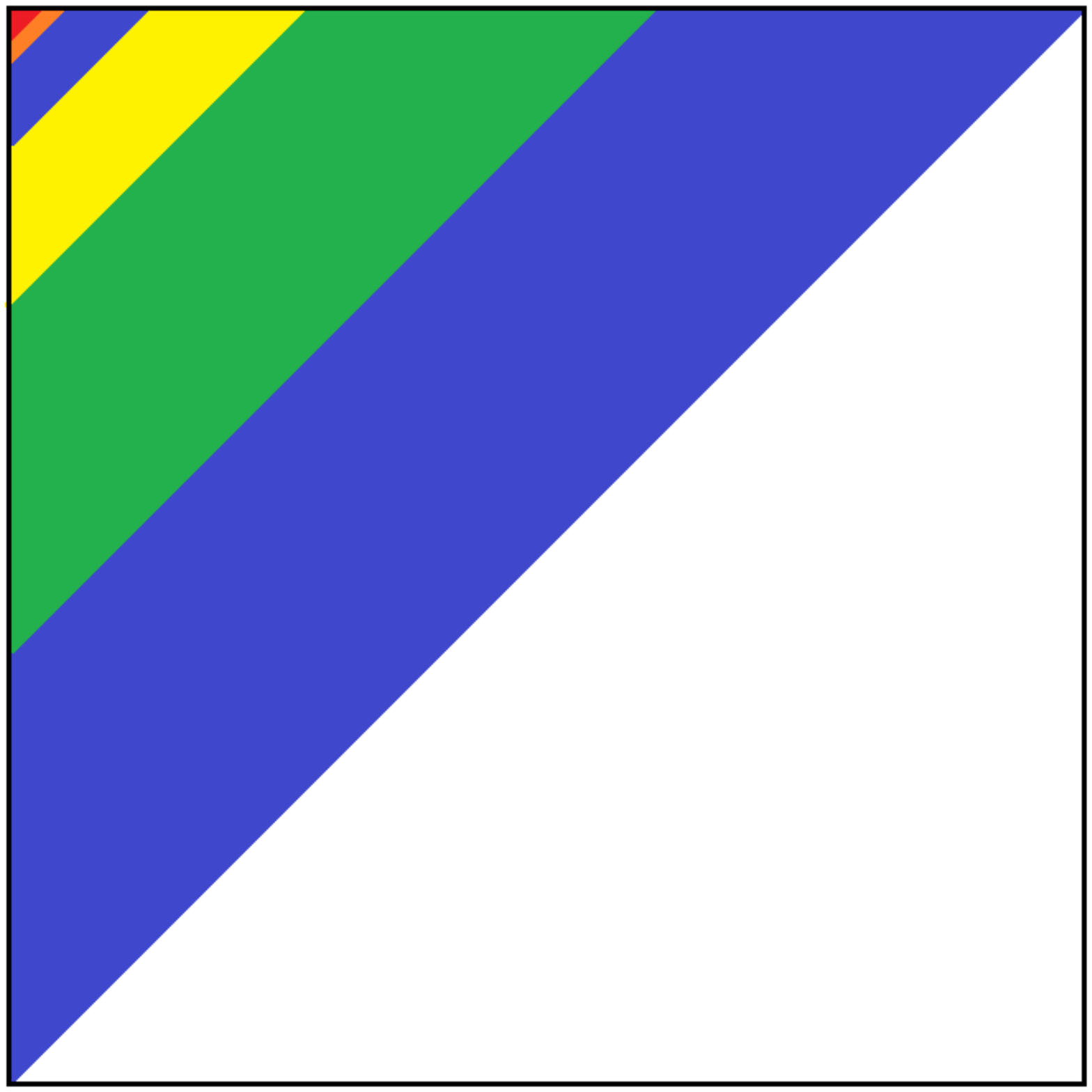}
\caption{This figure shows a series of nested bubble nucleations of vacua with different colors. The probabilities in the upper left corner are dictated by the dominant eigenvector.}
\label{f4}
\end{center}
\end{figure}

One possible loophole is that the observer in the red vacuum can observe the initial (blue) color by properties of the tunneling process from blue  to red. To see that this is not correct, consider Figure \ref{f4}
This figure shows a series of nested bubble nucleations that takes place in the causal patch of the red observer. The probability distributions of colors is exactly that given by the rate equations and is therefor controlled at late time by the dominant eigenvector, not the initial condition. The dominant eigenvector is associated with the asymptotic late-time evolution and has no memory of the initial (blue) starting point.

\bigskip

 \section{Breaking of Conformal Symmetry and the Persistence of Memory}

There is another related argument for the existence and relevance of an initial condition. It is connected with the remarkable phenomenon called ``persistence of memory" (PoM) by Garriga, Guth, and Vilenkin \cite{Garriga:2006hw}. PoM describes the the breaking of the de Sitter symmetry-group $O(4,1)$ due to bubble collisions.  The action of this group on the future boundary generates the conformal group in $3$-dimensions. PoM also breaks the conformal symmetry.

According to the authors of \cite{Garriga:2006hw} the persistence of memory is an unavoidable  observable effect of  initial conditions imposed on some initial surface.
However, although not stated in \cite{Garriga:2006hw}, the argument as given crucially depends on the existence of terminal vacua\footnote{In the toy model of \cite{Garriga:2006hw} the  decay of the descendant bubbles are not allowed to recycle to the false vacuum. In other words they are treated as terminal. }.
In the tree model of \cite{Harlow:2011az}\cite{Susskind:2012pp}  if there are no terminals then the equilibrium between so-called up and down transitions  leads to a conformally invariant fixed point\footnote{This is rigorously proved in the tree model but not in a more general context. }. The same thing
(the existence of a conformal fixed-point) is true in the Freivogel-Kleban model which does not have terminals but does allow bubble collisions \cite{Freivogel:2009rf}.

The existence of terminals  does in fact break de Sitter or conformal invariance, as well as leading to an arrow-of-time \cite{Susskind:2012pp}, and a preferred time-slicing. The question however, is whether it implies a finite past. The breaking of de Sitter symmetry in the tree model  is a feature of the asymptotic late-time fractal-flow attractor.
PoM is probably best thought of as a spontaneous symmetry breaking in which  a frame is selected in the eternal past, and then frozen in by the dynamics of terminal decays.    It does not imply that the multiverse is finitely old.

What I have said about PoM requiring terminals does not apply to certain non-observable effects having to do with correlation functions on scales much bigger than horizon scales, even without terminal vacua. For example a wavy initial surface would permanently imprint itself on correlation functions on global scales but these effects are unobservable\footnote{I thank Matt Kleban and Ben Freivogel for a discussion of this point.}.

\section{The Causal Patch Viewpoint}

My arguments have been in the context of the global or multiversal view of eternal inflation. Bousso argues for a more local viewpoint that focuses on a the evolution of a causal patch.  The causal patch starts in some state, or in some statistical ensemble of states.  If the landscape has a total of $e^{\cal{N}}$ colors, then after about ${\cal{N}}$ transitions it will reach a terminal vacuum. That is not very many and one would expect the vacua that occur along the way to be strongly biased by the initial condition.

One possible initial ensemble is the dominant eigenvector. In that case there is a duality between the global and local (causal patch) viewpoint. The argument is very simple. From the global viewpoint the  rate equation \ref{global rate} represents the global population of different vacua. But the equivalent local rate equation \ref{local rate} governs the probability along any causal patch. One can choose any initial ensemble and later observations will be biased by the choice. Thus an initial condition is necessary.

As Bousso was among the first to point out,
if the initial condition for the local equation is the dominant eigenvector then the local and global probabilities will agree. Thus one can reinterpret the theory in terms of a single causal patch and a particular initial condition: the dominant eigenvector. But one does not have to make this choice.

I think an analogy is helpful in distinguishing the various viewpoints.
Let's consider three models of an ordinary  universe with planets in it.  In the first model---the one-world-model---a single planet was created a finite time from the initial creation event.  The planet lives for a time and then dies. Any observations that take place on the planet take place a finite time after creation. If the people on the planet have a theory of the creation event they can confirm it by observation.

The problem with this model is that the fine-turnings needed for life are unexplained. Using anthropic reasoning would be tantamount to invoking a miracle.

The second model is the way things really happened. In the many-world-model (this has nothing to do with the many-worlds interpretation of quantum mechanics) many planets, of which there are a large variety of types, were independently created, more or less at the same time after the Big-Bang. The planets live for a time and then die. Once again, if a planet happens to have observers, their observations take place a finite time after the Big-Bang. The observers can confirm their theory of creation.

The big difference between the one-world-model and the many-world-model is that anthropic explanations make sense in the latter.

In the third model planets reproduce. The universe begins with a single planet which reproduces after a time and then dies. The process continues indefinitely so that the planetary population increases exponentially. Mutations can occur so that a landscape of planetary types is populated. For the same reasons that I discussed earlier, the average time from creation, that an observation takes place, is infinite. For all but a set of measure zero, the probability that a planet is of type $n$ is completely insensitive to the initial condition. It is determined by the dominant eigenvector of a rate equation, but only because the universe is very old when it is observed.

The reproducing-world-model is of course most like the usual picture of eternal inflation.
The use of anthropic reasoning is complicated by the same measure ambiguities that plague eternal inflation.

 I believe that what Bousso is contemplating (on Mondays, Wednesdays, and Fridays) is more like the  many-world-model in which initial conditions matter, but anthropic reasoning makes sense.  What, exactly such a model has to do with eternal inflation is not clear to me.

Lastly Bousso considers a kind of hybrid of the many-world and reproducing-world models. It is the many-world model, but with the initial condition being defined by the dominant eigenvector of the reproducing model. This model is contrived so that the local and global models are observationally indistinguishable. The observers would be able to tell that the ensemble of vacua that gave birth to their universe was the dominant eigenvector but they would have no way of knowing if they live in an infinitely old multiverse or in a young universe with an initial state set by the dominant eigenvector. There is a surface resemblance between this idea and the hypothesis that the earth was created six-thousand years ago with the dinosaur bones all in place, but that may be selling the hybrid model short. While it is contrived, it does have the same predictive power as the global theory, and in some ways is easier to use.

\section{Observational Consequences of the Dominant Eigenvector Hypothesis}

Let me come now to the question in the second half of the title: ``And What if It Is?"

The predictive power of the hypothesis of a very old multiverse lies in the implication that
 prior probability distribution  is given by the dominant eigenvector \cite{Garriga:2005av}. The dominant eigenvector provides a statistical bias about where our vacuum lies on the string-theory landscape. As Douglas has speculated \cite{Douglas}, the dominant eigenvector hypothesis can potentially be a basis for  phenomenology as well as an efficient seach in the landscape.   
 
  The phenomenological applications  of the dominant eigenvector hypothesis make use of a conjecture about the landscape. Because transition rates are doubly exponentially small\footnote{Transition rates are inverse exponentials of tunneling actions which themselves are often very large.} the ratio of decay rates are typically very far from $1$. A likely consequence of this highly staggered landscape is that the dominant eigenvector is strongly concentrated near one particluar vacuum---in Douglas' terminology the master vacuum\footnote{Previous authors used the term dominant vacuum. To avoid confusion with the dominant eigenvector I will use Douglas' terminology.}---whose color I will label $d$. The master vacuum is the longest lived vacuum on the landscape. To define it we consider the total decay rate of the vacuum $m$ to all vacua including terminals.
  \be
  \Gamma_m = \sum_n \gamma_{nm} + \gamma_m
  \ee
  
 Given that the decay rates are doubly exponential, the ratios of these rates
 are far from one, and one expects some particular vacuum to have a much smaller $\Gamma$ than any other. That vacuum is the master vacuum $d$. 
 
 The dominant eigenvector is clustered around $d$ in the following sense. The landscape can be viewed as a lattice whose points are connected if there is a direct \cdl \ instanton connecting them. The eigenvalues and eigenvectors of the rate equation are analogs of  the wave functions and energy levels of a fake hopping Hamiltonian on the lattice. The potential energy of a lattice site is $\Gamma_m$ and the kinetic (hopping) term is minus the symmetric matrix 
 \be
 t_{mn} = e^{-S_m+S_n}\gamma_{mn} = e^{-S_n+S_m}\gamma_{nm}
 \ee
  (see \cite{Harlow:2011az}).
  The dominant eigenvector can be thought of as a bound state localized around the lowest potential energy state $d.$ The wave functions falls off fast with the graph-distance from $d$ on the lattice.

Where should we look for the dominant vacuum? The likely answer is among the vacua with extremely small supersymmetry breaking scale, and small cosmological constant; since supersymmetric vacua are exactly stable. The master vacuum may or may not be the vacuum of absolutely lowest cosmological constant. Even if it is, the dominant eigenvector will not particularly resemble the Hartle-Hawking vacuum weighted by $e^{S_m}.$

 At first sight it would seem attractive to predict that we live in the master vacuum. However, that is not feasible---such a highly supersymmetric state would not be anthropically viable. Both for anthropic and observational reasons a period of slow-roll inflation is necessary, and to get to the slow-roll plateau the local universe must have tunneled from a nearby ancestor state. Thus one should look for that ancestor state which allows tunneling to the inflationary plateau, and which has the largest projection onto the dominant eigenvector. 
 
 Douglas argues that the neighborhood of the master vacuum may be very rich with enough vacua to provide numerous anthropically allowable cases, and also candidates for ancestor vacua to decay to the anthropic vacua. Given the proximity to the almost supersymmetric vacuum $d$, one may expect that the vacuum favored by the dominant eigenvector hypothesis will also have a relatively low supersymmetry breaking scale. The same argument may also  favor a low inflationary scale.

The purpose of this paper is not to do phenomenology but to lay a foundation for the use of the dominant eigenvector. The main point is that the dominant eigenvector has nothing to do with initial conditions but instead has to do with an asymptotically late-time attractor called a fractal-flow. Apart from the possiblility of a conspiracy, the attractor will only dominate if the multiverse has undergone 
an enormous number of transitions. This number is far larger than $\cal{N},$ the expected number of transitions that a causal patch can survive before decaying to a terminal. This is not an obstacle if the global picture of eternal inflation is correct, since global measures imply that the probability for any finite age is zero.

\section*{Acknowledgements}

I am grateful to Alex Vilenkin, Alan Guth, and Raphael Bousso for an extended email conversation which helped clarify the issues. I also thank Michael Douglas for a preview of \cite{Douglas} before publication, and Steve Shenker for explaining the importance of the master vacuum.

This work is supported by the Stanford Institute for Theoretical Physics and NSF Grant 0756174.

  \end{document}